\documentclass[prl,twocolumn,showpacs,superscriptaddress]{revtex4}

\usepackage{graphicx}
\usepackage{hyperref}
\usepackage{amssymb,amsmath}
\usepackage{epstopdf}
\usepackage{ulem}

\begin{document}

\title{Experimental Investigation of the Transition between Autler-Townes Splitting \\
and Electromagnetically-Induced Transparency}

\author{L. Giner}
\affiliation{Laboratoire Kastler Brossel, Universit\'{e}
Pierre et Marie Curie, Ecole Normale Sup\'{e}rieure, CNRS, 4 place
Jussieu, 75252 Paris Cedex 05, France}
\author{L. Veissier}
\affiliation{Laboratoire Kastler Brossel, Universit\'{e}
Pierre et Marie Curie, Ecole Normale Sup\'{e}rieure, CNRS, 4 place
Jussieu, 75252 Paris Cedex 05, France}
\author{B. Sparkes}
\affiliation{Centre for Quantum Computation and Communication
Technology, Australian National University, Canberra, ACT 0200,
Australia}
\author{A. S. Sheremet}
\affiliation{Laboratoire Kastler Brossel, Universit\'{e}
Pierre et Marie Curie, Ecole Normale Sup\'{e}rieure, CNRS, 4 place
Jussieu, 75252 Paris Cedex 05, France}
\affiliation{Department of Theoretical Physics, State
Polytechnic University, 195251, St.-Petersburg, Russia}
\author{A. Nicolas}
\affiliation{Laboratoire Kastler Brossel, Universit\'{e}
Pierre et Marie Curie, Ecole Normale Sup\'{e}rieure, CNRS, 4 place
Jussieu, 75252 Paris Cedex 05, France}
\author{O. S. Mishina}
\affiliation{Universit\"{a}t des Saarlandes, Theoretische
Physik, 66123 Saarbr\"{u}cken Germany}
\author{M. Scherman}
\affiliation{Laboratoire Kastler Brossel, Universit\'{e}
Pierre et Marie Curie, Ecole Normale Sup\'{e}rieure, CNRS, 4 place
Jussieu, 75252 Paris Cedex 05, France}
\affiliation{ONERA, The French Aerospace Lab, Chemin de la
Huni\`{e}re, 91761 Palaiseau, France}
\author{S. Burks}
\affiliation{Laboratoire Kastler Brossel, Universit\'{e}
Pierre et Marie Curie, Ecole Normale Sup\'{e}rieure, CNRS, 4 place
Jussieu, 75252 Paris Cedex 05, France}
\author{I. Shomroni}
\affiliation{Chemical Physics Department, Faculty of
Chemistry, Weizmann Institute of Science, Rehovot, Isra\"{e}l}
\author{D. V. Kupriyanov}
\affiliation{Department of Theoretical Physics, State
Polytechnic University, 195251, St.-Petersburg, Russia}
\author{P. K. Lam}
\affiliation{Centre for Quantum Computation and Communication
Technology, Australian National University, Canberra, ACT 0200,
Australia}
\author{E. Giacobino}
\affiliation{Laboratoire Kastler Brossel, Universit\'{e}
Pierre et Marie Curie, Ecole Normale Sup\'{e}rieure, CNRS, 4 place
Jussieu, 75252 Paris Cedex 05, France}
\author{J. Laurat}\email[]{julien.laurat@upmc.fr}
\affiliation{Laboratoire Kastler Brossel, Universit\'{e}
Pierre et Marie Curie, Ecole Normale Sup\'{e}rieure, CNRS, 4 place
Jussieu, 75252 Paris Cedex 05, France}

\begin{abstract}
Two phenomena can affect the transmission of a weak signal field
through an absorbing medium in the presence of a strong additional
field: electromagnetically induced transparency (EIT) and
Autler-Townes splitting (ATS). Being able to discriminate between
the two is important for various applications. Here we present an
experimental investigation into a method that allows for such a
disambiguation as proposed in [Phys. Rev. Lett. \textbf{107}, 163604
(2011)]. We apply the proposed test based on Akaike's information
criterion to a coherently driven ensemble of cold cesium atoms and
find a good agreement with theoretical predictions, therefore
demonstrating the suitability of the method. Additionally, our
results demonstrate that the value of the Rabi frequency for the
ATS/EIT model transition in such a system depends on the level
structure and on the residual inhomogeneous broadening.
\end{abstract}

\date{\today}
\maketitle

Fine engineering of interactions between light and matter is
critical for various purposes, including information processing and
high-precision metrology. For more than two decades, coherent
effects leading to quantum interference in the amplitudes of optical
transitions have been widely studied in atomic media, opening the
way to controlled modifications of their optical properties
\cite{Fleischhauer}. More specifically, such processes as coherent
population trapping \cite{Alzetta1976,Arimondo}  or
electromagnetically induced transparency (EIT)
\cite{Harris1990,Harris1991,Marangos} allow one to take advantage of
the modification of an atomic system by a so-called control field to
change the transmission characteristics of a probe field. These
features are especially important for the implementation of optical
quantum memories \cite{Tittel} relying on dynamic EIT \cite{Hau}, or
for coherent driving of a great variety of systems, ranging from
superconducting circuits \cite{Kelly} to nanoscale optomechanics
\cite{Safavi}.

However, if in general the transparency of an initially absorbing
medium for a probe field is increased by the presence of a control
field, two very different processes can be invoked to explain it in
a $\Lambda$-type configuation. One of them is a quantum Fano
interference between two paths in a three-level system
\cite{Fano1961}, which occurs even at very low control intensity and
gives rise to EIT \cite{Harris97}. The other one is the appearance
of two dressed states in the excited level at large control
intensity, corresponding to the Autler-Townes splitting (ATS)
\cite{Autler1955,Cohen77,Cohen}. Discerning whether a transparency
feature observed in an absorption profile is the signature of EIT or
ATS is therefore crucial \cite{Anisimov,Salloum,Zhang}. A recent
paper by P.M. Anisimov, J.P. Dowling and B.C. Sanders
\cite{Sanders2011} introduced a versatile and quantitative test to
discriminate between these two phenomena.

In this paper, we report an experimental study of the proposed
witness, relying on a detailed analysis of the absorption profile of
a probe field in an atomic ensemble in the presence of a control
field. In order to analyze the quantum interferences in detail and
avoid any inhomogeneous broadening, our study is performed with an
ensemble of cold cesium atoms in a well-controlled magnetic
environment. We show that the general behavior is in agreement with
Ref. \cite{Sanders2011}, but we identify some quantitative
differences. We finally interpret the characteristics of the EIT to
ATS model transition by taking into account the multilevel structure of
the atomic system and some residual inhomogeneous broadening.

\begin{figure*}[t]
\includegraphics[width=1.85\columnwidth]{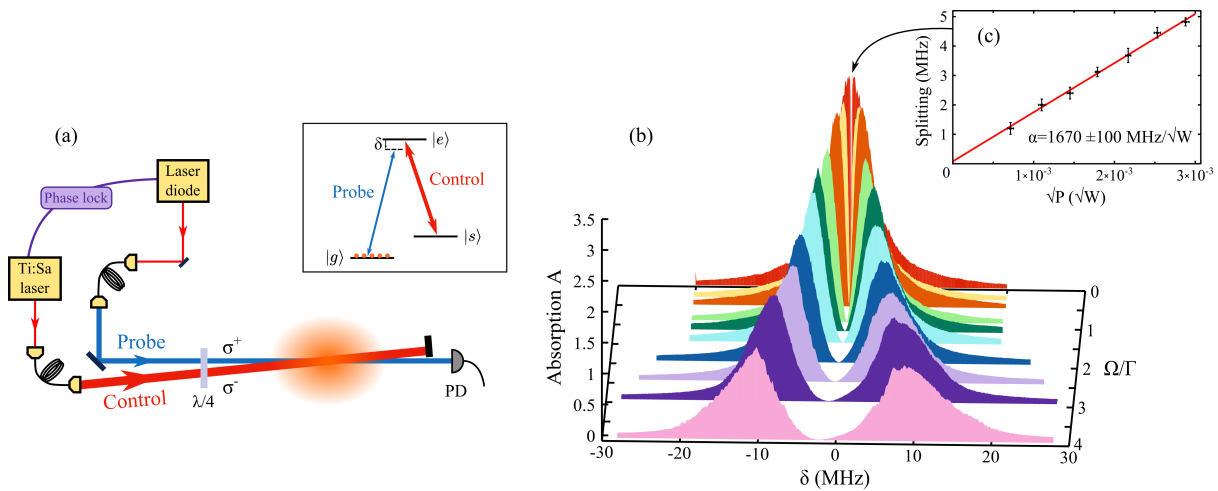}
\caption{(color online). Absorption in a $\Lambda$-type system. (a)
Experimental setup: a weak probe beam and a control beam travel
through a cloud of cold cesium atoms. Atoms are initially in the
ground state $|g\rangle$. The probe field is close to the
$|g\rangle\rightarrow |e\rangle$ transition (detuning $\delta$)
while the control field drives the $|s\rangle \rightarrow |e\rangle$
transition on resonance; PD : high-gain photodiode. (b) Absorption
profiles are displayed as a function of the detuning $\delta$ for a control Rabi frequency $\Omega$ between $0.1
\Gamma$ and $4 \Gamma$, where $\Gamma$ is the natural linewidth. (c)
Experimental splitting as a function of $\sqrt{P}$, where $P$ is the
measured control power, fitted by a linear function.}\label{exp}
\end{figure*}

The experimental setup is illustrated in Fig. \ref{exp}(a). The
optically thick atomic ensemble is obtained from cold cesium atoms
in a magneto-optical trap (MOT). The three-level $\Lambda$ system
involves the two ground states, $|g\rangle=|6S_{1/2},F=3\rangle$ and
$|s\rangle=|6S_{1/2},F=4\rangle$, and one excited state
$|e\rangle=|6P_{3/2},F=4\rangle$. The control field is resonant with
the $|s\rangle$ to $|e\rangle$ transition, while the probe field is
scanned around the $|g\rangle$ to $|e\rangle$ transition, with a
detuning $\delta$ from resonance.

Each run of the experiment involves a period for the cold atomic
cloud to build up and a period for measurement. This sequence is
repeated every 25 ms and controlled with a FPGA board. After the
build-up of the cloud in the MOT, the current in the coils
generating the trapping magnetic field and then the MOT trapping
beams are switched off. In order to transfer the atoms from the
$|s\rangle$ to the $|g\rangle$ ground state, the MOT is illuminated
with a $\sigma^{-}$-polarized 1 ms-long depump pulse with a power of
900 $\mu$W and resonant with the $|s\rangle$ to $|e\rangle$
transition. After this preparation stage, the optical depth at
resonance for atoms in $|s\rangle$ is zero within our experimental precision. The remaining spurious
magnetic fields have been canceled down to 5mG using a RF
spectroscopy technique.

The measurement period starts 3 ms after the extinction of the MOT
magnetic field. The atomic ensemble is illuminated with a 30 $\mu$s-long control
 pulse and a probe pulse lasting 15 $\mu$s is sent during this time. The probe field is emitted by an extended cavity
grating stabilized laser diode, whereas the control field is
generated by a Ti:Sapphire laser locked on resonance using saturated
absorption spectroscopy. The two lasers are phase-locked. The
control field is $\sigma^{-}$-polarized, with a 200 $\mu$m waist in
the MOT and a 2$^\circ$ angle relative to the direction of the probe
beam. The probe field is $\sigma^{+}$-polarized, with a waist of 50
$\mu$m and a power of 30 nW. To measure absorption profiles, the
probe beam frequency is swept over a few natural linewidths by
changing the locking frequency point. Its absorption is measured
with a high-gain photodiode. The optical depth in the $|g\rangle$
state is chosen to be around 3 to avoid any profile shape distortion
due to the limited dynamic range of the photodiode.

Figure \ref{exp}(b) gives the absorption of the probe field,
$A=\textrm{ln}(I_{\textrm{ref}}/I)$, as a function of its detuning
$\delta$ from resonance for different values of the control power
(0.1 to 200 $\mu$W), i.e. for different values of the control Rabi
frequency $\Omega$. The quantity $I_{\textrm{ref}}$, which gives the
transmission in the absence of atoms, is measured by sending an
additional probe pulse when all the atoms are still in the
$|s\rangle$ ground state. The Rabi frequency $\Omega$ of the control
field is changed from very weak values (at the back) to four times
the natural linewidth $\Gamma$ (at the front). Each profile results
from an averaging over twenty repetitions of the experiment. The
narrow transparency dip appearing for low Rabi frequencies gets
wider when the Rabi frequency increases, to finally give two
well-separated resonances corresponding to the two excited dressed
states.

Let us note that the Rabi frequency $\Omega$ is a linear function of
the electric field, and can be expressed as
$\Omega=\alpha\,\sqrt{P}$, with $P$ the power of the control field.
An effective value of $\Omega$ can be inferred from the experimental
splittings (i.e. the distance between the two maxima) observed in
the absorption profiles for low-power control field (Fig.
\ref{exp}(c)). For a three-level system this splitting
 is indeed equal to the Rabi frequency within a very good approximation for low decoherence
in the ground state \cite{Arimondo}. We find $\alpha=1670\pm 100
\,\textrm{MHz}/\sqrt{\textrm{W}}$.

We now turn to the detailed analysis of the absorption profiles. For
a three-level $\Lambda$ system, to first order in the probe electric
field, the atomic susceptibility on the probe transition for a
control field on resonance is given by \cite{Anisimov}:
\begin{eqnarray}\label{khi}
\chi(\delta)=-\frac{n_{g}
|d_{eg}|^{2}}{\mathbf{\hbar}\epsilon_{0}}\,
\frac{\delta+i\gamma_{gs}}{\delta^2-|\Omega_{0}|^2/4-\gamma_{eg}\gamma_{gs}+i\delta(\gamma_{eg}+\gamma_{gs})}
\end{eqnarray}
${n_{g}}$ stands for the atomic density in state $|g \rangle$ and
$d_{eg}$ denotes the electric dipole moment between $|e\rangle$ and
$|g\rangle$. Here, the Rabi frequency of the control field is
$\Omega_{0}=2\,|d_{es}|\varepsilon_c/\hbar$, with $\varepsilon_c$
the amplitude of the positive frequency part of the control field.
The optical coherence relaxation rate is $\gamma_{eg}=\Gamma/2$
where $\Gamma/2\pi=5.2 $ MHz. $\gamma_{gs}$ is the dephasing rate of
the ground state coherence, $\gamma_{gs}= 10^{-2}\Gamma$ in our
experimental case.

Depending on the value of the control Rabi frequency $\Omega_{0}$,
Eq. \ref{khi} can be rewritten in different ways
\cite{Anisimov,Salloum,Zhang}. For Rabi frequencies
$\Omega_{0}<\Omega_t=\gamma_{eg}-\gamma_{gs}$, the spectral poles of
the susceptibility are imaginary. Then, the linear absorption
$A\propto\textrm{Im}[\chi]$ can be expressed as the difference
between two Lorentzian profiles centered at zero frequency, a broad
one and a narrow one. For $\Omega_{0}>\Omega_t$, this decomposition
is not possible anymore. For large Rabi frequencies,
$\Omega_{0}\gg\Gamma$, Eq. \ref{khi} can be written as the sum of
two well separated Lorentzian profiles with similar widths. Absorption profiles for these two model can thus be written as:
\begin{eqnarray}
A\sb{\textrm{EIT}}&=\frac {C_{+}} {1 + (\delta -
\epsilon)^{2}/(\gamma_{+}^{2} /4)} - \frac {C_{-} } {1 +
\delta^{2}/(\gamma_{-}^{2} /4)} \label{EIT}\\
A\sb{\textrm{ATS}}&=\frac {C_{1}} {1 + (\delta +
\delta_{1})^{2}/(\gamma_{1}^{2}/4)} + \frac {C_{2}} {1 + (\delta -
\delta_{2})^{2}/(\gamma_{2}^{2}/4)} \label{ATS}
\end{eqnarray}
where $C_{+}$, $C_{-}$, $C_{1}$, $C_{2}$ are the amplitudes of the
Lorentzian curves, $\gamma_{+}$ and $\gamma_{-}$, $\gamma_{1}$ and
$\gamma_{2}$  are their widths, $\epsilon$, $\delta_{1}$ and
$\delta_{2}$ are their shifts from zero frequency. Equation
\ref{EIT} describes a Fano interference and corresponds to the EIT
model, while Eq. \ref{ATS} corresponds a strongly-driven regime with
a splitting of the excited state, i.e. ATS.

For a three-level system, the various parameters introduced in the
two above expressions can be calculated from Eq. \ref{khi}.
Conversely, in our experimental system, we use functions
$A\sb{\textrm{EIT}}$ and $A\sb{\textrm{ATS}}$ to fit the
experimental absorption curves, adjusting all the aforementioned
parameters. The test proposed in \cite{Sanders2011} aims at
determining which of these generic models is the most likely for
given experimental data.

\begin{figure}[t]
\includegraphics[width=0.95\columnwidth] {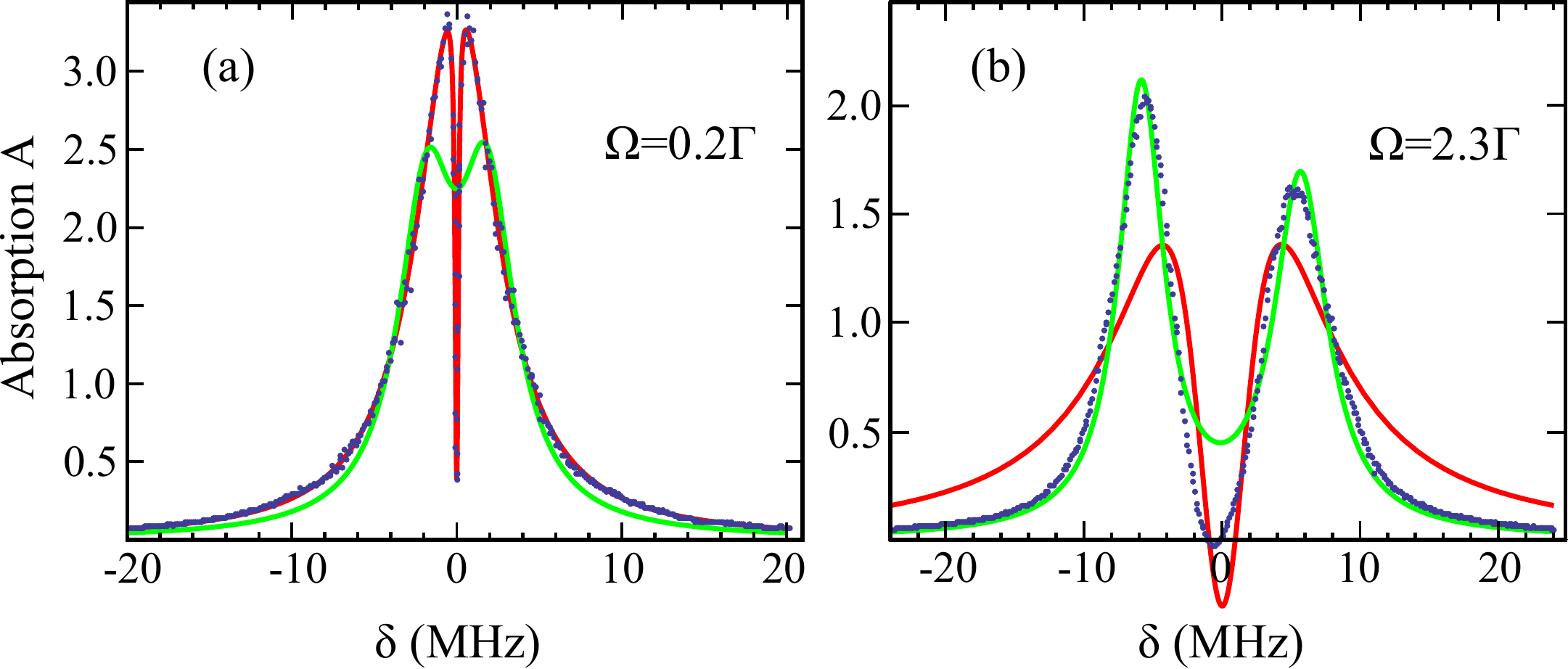}
\caption{(color online).  Absorption profiles and model fits for two values of the control Rabi frequency $\Omega$.
Experimental data (blue dots) are presented together with the best
fits of functions
$A\sb{\textrm{EIT}}(C_{+},C_{-},\epsilon,\gamma_{+},\gamma_{-})$
(red solid lines) and
$A\sb{\textrm{ATS}}(C_{1},C_{2},\delta_{1},\delta_{2},\gamma_{1},\gamma_{2})$
(green solid lines). Parameters $C_{+},C_{-},C_{1},C_{2}$ representing the
amplitudes of the absorption curves are in dimensionless units,
while the parameters $\epsilon,\gamma_{+},\gamma_{-}$ and
$\delta_{1},\delta_{2},\gamma_{1},\gamma_{2}$ representing
detunings and widths are in MHz. (a) $\Omega = 0.2
\Gamma$. In this case $A\sb{\textrm{EIT}}(3.52, 3.14, 1.45 \times
10^{-2},
 5.71, 0.239)$ fits the experimental
data much better than $A\sb{\textrm{ATS}}( 2.01, 2.04 ,1.84, 1.84,
4.14, 4.08)$. (b) $\Omega= 2.3 \Gamma$. Here $A\sb{\textrm{ATS}}(
2.05, 1.64, 5.86, 5.67, 3.94, 4.68)$ fits the data better than
$A\sb{\textrm{EIT}}(1.59 \times 10^{5}, 1.59 \times 10^{5}, 1.45
\times 10^{-7}, 8.17, 8.17)$.} \label{curves}
\end{figure}

Figure \ref{curves} shows the measured probe absorption as a
function of the detuning $\delta$ (blue dots) together with the fits
to $A\sb{\textrm{EIT}}$ (red curves) and $A\sb{\textrm{ATS}}$ (green
curves). A low value of the control Rabi frequency,
$\Omega=0.2\Gamma$ is shown in panel (a), and a larger one,
$\Omega=2.3\Gamma$ in panel (b). Let us note that for the EIT model
a detuning parameter $\epsilon$ was introduced between the atomic
line center and the EIT dip to account for a possible experimental
inaccuracy in the frequency locking reference of the lasers. For the
ATS model the parameters describing each Lorentzian curve are
independent of each other (contrary to what would be deduced from
Eq. \ref{khi}) in order to account for their experimentally
different widths and heights. These asymmetries are discussed below.
As expected, the EIT model fits better the low-power control field
region (panel (a)) while the ATS model fits better the strong-power
control field region (panel (b)).

\begin{figure}[t]
\includegraphics[width=.95\columnwidth]{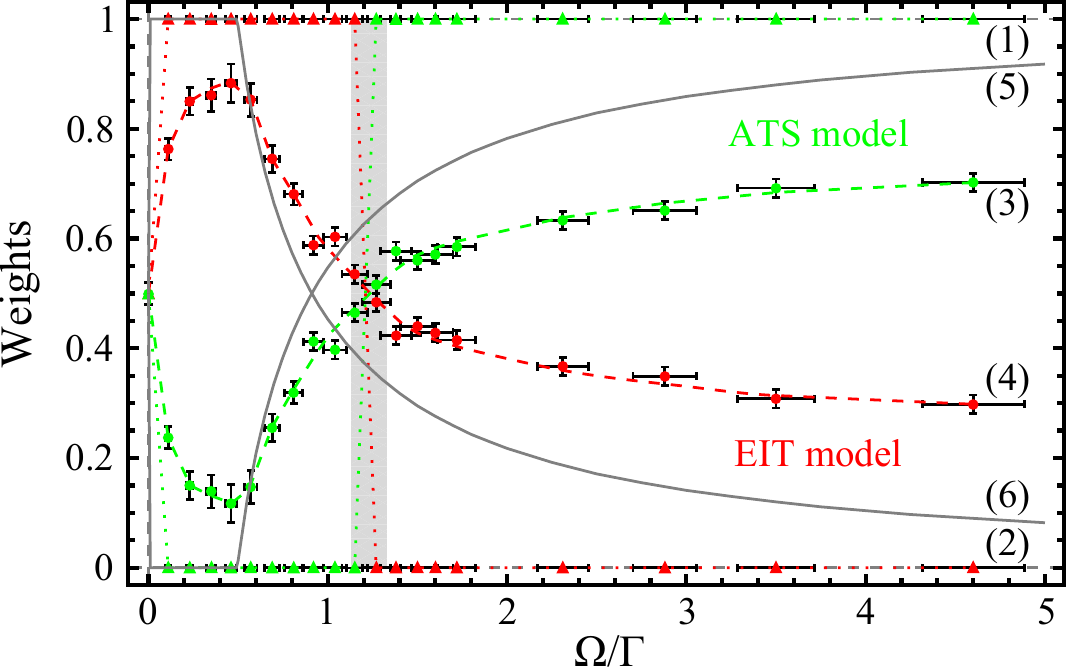}
\caption{(color online). Experimental Akaike weights $w_{j}$ as a
function of the Rabi frequency $\Omega$ for ATS model (green
triangles, curve (1)) and for EIT model (red triangles, curve (2)).
Experimental per-point weights $\overline{w}_{j}$ for ATS model
(green dots, curve (3)) and for EIT model (red dots, curve (4)). The
grey area indicates the EIT/ATS model transition. Error bars include the
uncertainty on the coefficient $\alpha$ and on the measured power. The
solid lines give the  theoretical per-point weights for a pure
3-level system (curves (5) and (6)).} \label{criterion}
\end{figure}

As proposed in \cite{Sanders2011}, in order to quantitatively test
the quality of these model fits, we then calculate the \textit{Akaike
information criterion} (AIC) \cite{Akaike}. This criterion, directly
provided by the function \textsc{NonLinearModelFit} in
\textsc{Mathematica}, is equal to $I_{j}=2 k - \textrm{ln}(L_{j})$
where $k$ is the number of parameters used and $L_{j}$ the maximum
of the likelihood function obtained from the considered model,
labeled with $j$ ($j=\textrm{EIT}$ or $\textrm{ATS}$). The relative
weights $w_{\textrm{EIT}}$ and $w_{\textrm{ATS}}$ that give the
relative probabilities of finding one of the two models can be
calculated from these quantities and are given by:
\begin{equation}
w_{\textrm{EIT}}=\dfrac{e^{-I_{\textrm{EIT}}/2}}{ e^{-I_{\textrm{EIT}}/2}+ e^{-I_{\textrm{ATS}}/2}},\,\,w_{\textrm{ATS}}=1-w_{\textrm{EIT}}.\nonumber
\end{equation}
These weights are plotted in Fig. \ref{criterion} (curves (1) and
(2)), as a function of the experimentally determined Rabi frequency.
They exhibit a binary behavior. They are close to 0 or 1 and there
is an abrupt transition from EIT model to ATS model.

We then investigate the second criterion proposed in Ref.
\cite{Sanders2011}, also based on Akaike's information criterion but
with a mean per-point weight $\overline{w}$. It can be obtained by
dividing $I_{j}$ by the number of experimental points \textrm{N}.
The weights for the EIT and the ATS models are now given
respectively by
\begin{equation}
\overline{w}_{\textrm{EIT}}=\dfrac{e^{-I_{\textrm{EIT}}/2N}}{ e^{-I_{\textrm{EIT}}/2N}+
e^{-I_{\textrm{ATS}}/2N}}, \,\, \overline{w}_{\textrm{ATS}}=1-\overline{w}_{\textrm{EIT}}.\nonumber
\end{equation}
The resulting curves are presented in Fig. \ref{criterion} (curves
(3) and (4)). Starting from a per-point weight equal to 0.5 for both
models in the absence of control field (the two models are equally
likely), the EIT model first dominates in the low Rabi frequency
region. Then the likelihood of the EIT model decreases and a
crossing is observed for the same value as for the previous
criterion. The ATS model then dominates for larger Rabi frequency,
as expected.

For the Akaike weights, as well as for the per-point weights, the
behavior is in good qualitative agreement with the predictions given
in Ref. \cite{Sanders2011} and with our simulations for a
three-level system. However, the transition between the two models
is obtained experimentally for $\Omega/\Gamma=1.23\pm 0.10$, while a
value of $\Omega/\Gamma=0.91$ is obtained for the per-point weights
of a pure three-level system, calculated for the same Rabi
frequencies, as shown in Fig. \ref{criterion} (solid grey lines).
Moreover, for large Rabi frequencies the per-point weights
corresponding to ATS and EIT saturate at 0.7 and 0.3 respectively
instead of going to 1 and 0 as in the theoretical three-level model.
For low Rabi frequencies, the shape of the curves also differs
significantly. These various features suggest that the system cannot
be described by a simple three-level model. Below, we proceed to
theoretical simulations including additional parameters that
influence the ATS/EIT model transition and the general shape of the
per-point weight curves.

\begin{figure}[htpb!]
\includegraphics[width=.9\columnwidth]{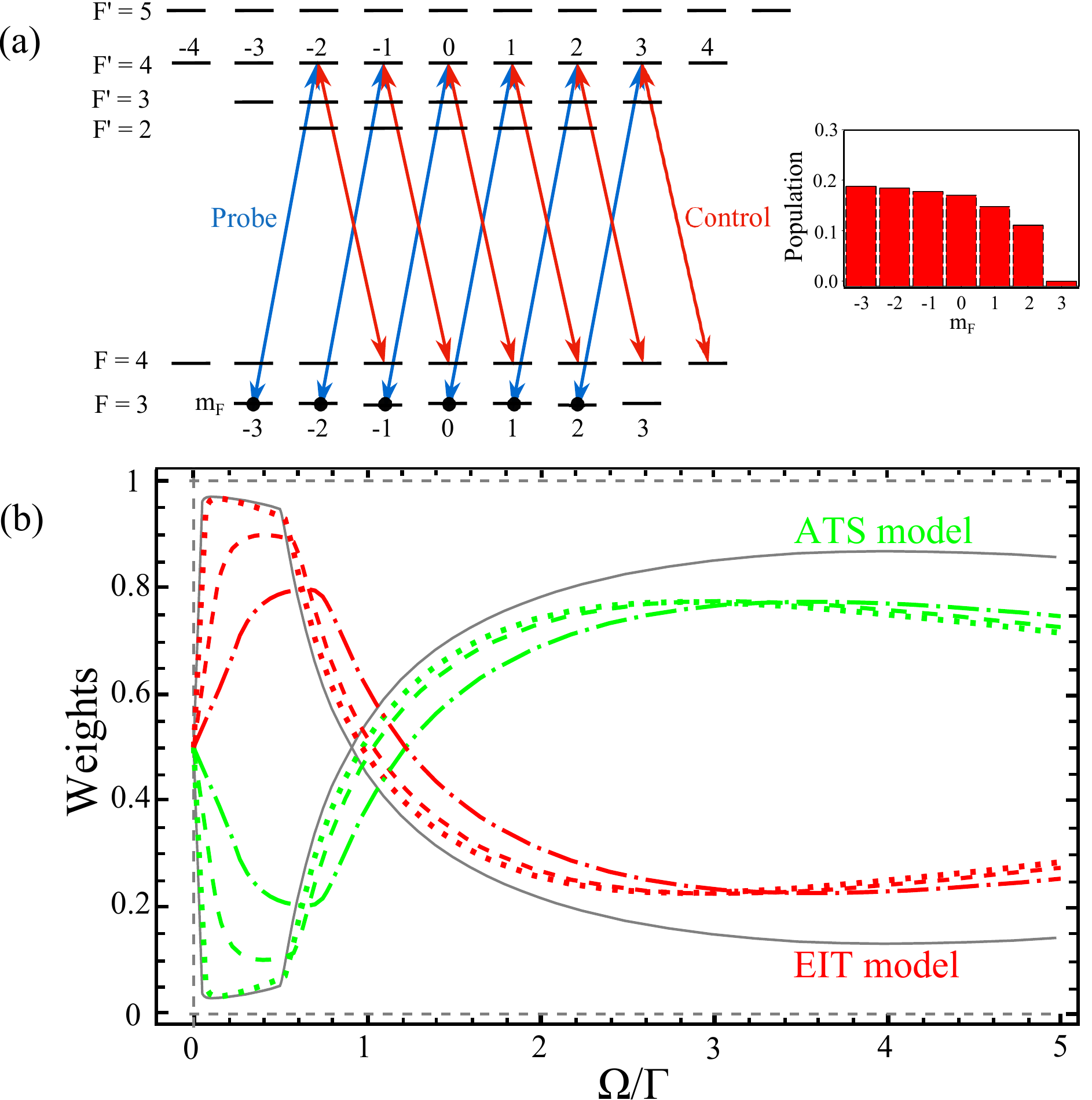}
\caption{(color online). Theoretical simulations with Zeeman
sublevels and Doppler broadening. (a) Level scheme for the Cs
$D_{2}$ line and the six three-level transitions involving Zeeman
sublevels; inset: population distribution. (b) Per-point weights for
a three-level system (grey solid lines), for a system involving
Zeeman sublevels (dotted lines), for a system involving Zeeman
sublevels and residual Doppler broadenings  $\Gamma_D/2\pi=0.6$ MHz
(dashed lines) and $\Gamma_D=1.3/2\pi$ MHz (dash-dotted
lines).}\label{fig4}
\end{figure}

First, we take into account the other hyperfine sublevels of the
$6P_{3/2}$ manifold, based on a previous theoretical model
\cite{Oxy2011,Michael2012}. We find that these contributions explain
the asymmetry between the two dressed-state resonances observed in
Fig. \ref{curves}(b) at large Rabi frequencies but they do not
significantly influence the per-point weight curves. The latter are
shown in Fig. \ref{fig4}(b) (solid lines), with a crossing point for
$\Omega/\Gamma=0.91$.

We then consider the effect of the Zeeman structure. Several Zeeman
sublevels are involved in each atomic level as shown in Fig.
\ref{fig4}(a). We have determined the atomic distribution in the
Zeeman sublevels from the optical pumping due to the depump field
(Fig. \ref{fig4}(a), inset). Since the control and probe fields have
opposite circular polarizations, we can consider that the atomic
scheme is a superposition of six independent $\Lambda$ subsystems
with different Rabi frequencies. The susceptibility is calculated as
the sum of the corresponding susceptibilities. The per-point weights
for theoretical absorption curves calculated from this model
(including the hyperfine structure) are shown in Fig. \ref{fig4}(b)
(dotted lines). For the horizontal axis, as the system does not have
a single Rabi frequency, we have used an effective Rabi frequency
obtained from the splitting between the maxima of the theoretical
absorption curves. The transition point is found for
$\Omega/\Gamma=0.98$, close to the value obtained for a three-level system. These simulations show that taking into account the
Zeeman sublevels does not lead to a large enough alteration of the
crossing point as compared to the three-level model. However a
significant change in the values of the per-point weights for large
Rabi frequencies is obtained for the model including the Zeeman
sublevels, and it is comparable to the experimental one.

We finally include a residual inhomogeneous Doppler broadening
$\Gamma_D$. By fitting the experimental absorption profile in the
absence of control field, we obtain $\Gamma_D/2\pi=0.6$ MHz. The
per-point weights for theoretical absorption curves including this
residual broadening are given in Fig. \ref{fig4}(b) (dashed lines).
The crossing point is found for a value $\Omega/\Gamma=1.05$, which
is in better agreement with the experimental value. The slightly
larger value of the experimental transition point is very likely to
be due to heating and additional broadening caused  by the control
laser. If we assume an inhomogeneous broadening $\Gamma_D/2\pi=1.3$
MHz (dash-dotted lines), the per-point weights for the theoretical
absorption curves cross each other for $\Omega/\Gamma=1.23$.
Moreover the shape of the curves including even a small Doppler
broadening agrees much better with the experimental results for the
low control Rabi frequency region. Thus, including in the model both
the Zeeman structure of the atomic system and a residual Doppler
broadening due to the finite temperature of the atoms allows us to
explain the observed experimental behaviour when the EIT-ATS
discrimination criterion is applied.

In summary, we have tested and analyzed in detail the transition
from the ATS model to the EIT model proposed in Ref.
\cite{Sanders2011} in a well controlled experimental situation. The
criteria have been calculated and give a consistent conclusion for
discerning between the two regions. The observed differences from
the three-level model have been interpreted by a refined model
taking into account the specific level structure and some residual
inhomogeneous broadening. This study confirms the sensitivity of the
proposed test to the specific properties of the medium and opens the
way to a new tool for characterizing complex systems involving
coherent processes.
\\

\acknowledgements This work was supported by the CHIST-ERA ERA-NET
(QScale project), by the Australian Research Council Centre of
Excellence for Quantum Computation and Communication Technology
(CE110001027), and by the CNRS-RFBR collaboration (CNRS 6054 and
RFBR P2-02-91056). A.S. acknowledges the support from the Foundation
"Dynasty" and O.S.M. from the Ile-de-France program IFRAF. J.L. is a
member of the Institut Universitaire de France.

\end{document}